\documentstyle[11pt]{article}

\begin{document}

\noindent
{\large \it Yukawa Institute, Kyoto} \hspace{8.7cm}{YITP-95-3}

\rightline{September 1995}
%\rightline{hep-ph/9509...}

\vskip 3cm

\begin{center}

{\LARGE \bf
Solar Neutrinos: Expecting 1996
}\footnote{Talk given at the International workshop
Particle Physics: Present and Future, Valencia, June 1995}
\\

\vskip 2cm

{\Large A. Yu. Smirnov}\\

\vskip 1.5mm

{\it Yukawa Institute for Theoretical Physics, Kyoto University,\\
Kyoto 606-01, Japan, and \\
International Centre for Theoretical Physics,\\
I-34100, Trieste, Italy}\\

\vskip 2cm

{\Large Abstract}

\end{center}

These are remarks (mainly on the solar neutrinos) written
in anticipation of 1996 - the year
which can be crucial for the neutrino physics.
Recent results on solar neutrinos are discussed.
The topics include: (solar) model independent
approach to the solar neutrino problem, status of
different solutions of the problem,
standard and non-standard scenarios of the lepton mixing,
light singlet fermions and the neutrino phenomenology.

%%%%%%%%%%%%%%%%%%%%%%%%%%%%%%%%%%%%%%%%%%%%%%%%%%%%%%%%%%%%%%%%%%%%%

\newpage

\section{Introduction}

According to time schedule~\cite{sup}
in January - March '96
stainless tank placed in the Kamioka mine will be filled
by 50.000 tons of water and in April 96 the
SuperKamiokande -- the  detector of new generation of the
underground experiments -- will start to take data.
Later  Sudbury detector~\cite{sno} will
to be ``on the line".
CHORUS~\cite{chor} is expected to announce  first results.
CHOOZ~\cite{choo} - the first long base line reactor experiment
will begin to operate soon.

The results from these experiments may resolve
neutrino puzzles we are discussing now.
In any case they  will have strong
impact not only on the neutrino physics
and astrophysics but also on particle
physics as whole.

Taking this into account it may be wise to abstain from
further  theoretical speculations and
just wait (laying a bet) for new experimental
results. On the other hand it is a good time to summarize what we
have learned
and to understand what is our starting point before new
results will arrive. It is a good time to formulate
{\it a priori}  criteria according to which we will make the
conclusions in future.

\section{Experiment and Theory}

\subsection{Experiment}

1. Results from all solar neutrino experiments~\cite{hom,kam,sag,gal}
 are stable.
There are small changes of the average signals within $1 \sigma$:
Homestake result has increased by $1 \sigma$~\cite{hom}, Kamiokande
flux decreased by statistical $1 \sigma$ ~\cite{kam}, the change
of GALLEX result is even smaller~\cite{gal}.

2. The error bars slowly decrease indicating that
even Gallium experiments become ``old".
When experiments become old it is a proper  moment to speak
about time variations. Search for time variations is
one of justifications to continue  the experiments.
Recent analysis of Gallium data shows very good agreement
with constant original flux and  even admission of  different time
variations  practically does not improve the
goodness of the fit~\cite{gal}.

Another justification is accuracy.
In fact, one can imagine a situation when, even  50\%
decrease of error bars in Gallium experiment could be decisive
for the problem.

Kamiokande does not see any time variations
(day-night, seasonal, anticorrelation with
solar activity etc..) at least with more than 30\% amplitude.
Of course, an experienced  eye can find  about 4 years period
``wave" (especially if one removes systematical errors).

During last 5 years Homestake data did not show neither
correlations nor anticorrelations with sunspot numbers
(and in this sense there is a good agreement with
Kamiokande negative result).
The confidence level of the anticorrelation
during all the period of observation is
approximately the same as in  1980 ~\cite{stan}.
That is  the effects is essentially due
to very low signal during  one year:  1979 - 1980. This was
the year of change of the magnetic field polarity in the Sun.
(Accidental coincidence?  This  can be check
in the year 2002).
Runs with high counting rate in the period of low activity
1986 - 1987 have rather peculiar statistical
distribution~\cite{smi}.

Anyway, SuperKamiokande will continue the Kamiokande job.
Iodine experiment can probably substitute Chlorine  monitoring,
and it will be certainly worthwhile to have Gallium
experiment working during the operation of SuperKamiokande
and SNO (just in case).

3. It is difficult to overestimate the importance
of the result  of GALLEX experiment with $^{51} Cr$
source~\cite{cro}. It gives not only the overall check of
the solar  experiment
and especially, the efficiency of detection
in the important energy region of the
$^7 Be$ neutrinos. It gives the credit to  radiochemical
method, and therefore the additional credit to the Homestake result.

4. One more result on the solar neutrinos: recently new
bound on the antineutrino flux from the Sun
$\Phi(\bar{\nu}_e)$ has been published
by LVD collaboration~\cite{lvd}
\begin{equation}
\Phi(\bar{\nu}_e) = (2 - 4) \cdot 10^{-3} cm^{-2}s^{-1},
\label{eq:ant}
\end{equation}
i.e. $\Phi(\bar{\nu}_e)/\Phi_B^{SSM} = 0.3 - 0.6$ \% ,
where $\Phi_B^{SSM}$ is the boron neutrino flux according to the SSM.
The bound has an important implication to the
spin-flavor precession of the solar neutrinos.

\subsection{Theory}

Standard Solar Model (SSM) predictions where updated recently
in a number of aspects.  New values of input parameters
like  nuclear cross sections, solar age, abundance of the elements,
radiative opacities etc.. are used. The
uncertainties related to the pre-Main
sequence evolution, to the depth of the convection zone etc.,  were studied
(see~\cite{bah} for review).
The most important changes are related to
taking in to account the diffusion of heavy elements
(C, N, O, ..., Fe, Ni ... ). The diffusion leads to
increase of the opacity in the
radiative zone, and consequently,
in increase of the central temperature of the Sun by
about 1.1 \%. As a result the boron and the beryllium neutrino
fluxes increase by 17 \% and 6 \%  correspondingly.

The solution of the  $^7 Li$-problem
- one of the origin of  doubt in the reliability of
the SSM itself and its predictions of the neutrino fluxes -
probably has been  found.  The reason of the Li- deficit
(whose surface concentration is about 200
times smaller than expected)
could be  in  plasma physics~\cite{li}.
It has been  pointed out that in the effective screening
potential of the electron
cloud the nuclei become more transparent to each other.
This increases
the effective energy of collisions by 600 - 700 eV which leads to
decrease of  temperature of the $^7 Li$  burning.
Also diffusion becomes more efficient.

\section{Experiment without Theory}

\subsection{Solar neutrino problem without solar model}

In spite of serious  progress in the solar modelling and  very good
agreement of SSM and helioseismological data, some predicted solar
neutrino fluxes  still have rather large uncertainties.  Mainly,
they are related to the nuclear cross-sections (first of all, for  the
reaction  $p + ^7 Be \rightarrow ^8 B + \gamma$)
and probably to some plasma effects
which have not yet been properly taken into account~\cite{plas}. \\

These uncertainties will hardly be fixed before new
experiments on solar neutrinos  start to operate.
In this connection  the approach
to the problem has been elaborated
which does not use the
absolute values of neutrino fluxes from the SSM.
Main points of the approach  which can be called
``solar neutrino problem without solar model"
are the following
{}~\cite{bar} - \cite{bfl}:

1. Only general notion is used about the
solar neutrinos: the composition and  the energy
spectra of
components, but not the absolute values of fluxes. These  absolute
values are considered as {\it free parameters to be found from the
solar neutrino experiments}.
In particular, the boron neutrino flux can be
represented  as
\begin{equation}
\Phi_B = f_B \cdot \Phi_B^{SSM},
\label{eq:bor}
\end{equation}
where
$f_B$ is free parameter, and $\Phi_B^{SSM}$ is the flux in the
reference SSM e.g.~\cite{bp}.
Similarly, the parameters $f_i$ $(i = Be, pp, NO)$ for other important
fluxes can be introduced.\\
2. The data from different experiments are confronted immediately.\\
3. The  solar neutrino fluxes are normalized on
the solar luminosity (the normalization
in based on  the condition of thermal
equilibrium of the Sun).\\

Already  existing  data allow one  (i) to formulate the
problem in practically model independent way, (ii) to restrict not
only neutrino parameters but also original neutrino fluxes.
In future precision  solar neutrino data will be used to get the
information on solar model~\cite{hala2}. Thus
we will turn to the original
proposal to study  the interior of the Sun by neutrinos.\\

There are two key points in the  analysis
of present experimental situation.

 {\it Kamiokande versus  Homestake}~\cite{bar} - \cite{degl}.
Suppose first that neutrino flux consists of $\nu_e$ only. Then
boron neutrino flux
measured by Kamiokande gives the contribution to  Ar-production
rate $Q_{Ar,B} = 3.00 \pm 0.45$ SNU
which exceeds the total signal observed
by Homestake: $Q_{Ar}^{obs} = 2.55 \pm 0.25$ SNU.
This means that the contributions  of all
other fluxes to $Q_{Ar}$, and in particular, of
Beryllium neutrinos should be strongly suppressed.

{\it Gallium Experiment Results versus Solar
Luminosity}~\cite{spi,cas,dar,ber}.
The luminosity of the Sun
allows one to estimate the
pp- neutrino flux, and consequently its contribution to
Ge-production rate: $Q_{Ge, pp} \approx 71$ SNU. This value plus small
($\sim 5$ SNU) contribution of boron neutrinos coincides
with total signal observed by GALLEX. Consequently, gallium
results can be reproduced if the beryllium neutrino flux
as well as all other fluxes
of the intermediate energies are strongly suppressed.

Thus both these points indicate on strong suppression of the $^7$Be-
neutrino flux (if there is no neutrino conversion).
Statistical analysis gives
$f_{Be} < 0.3$ (2$\sigma$) (for more detail
see~\cite{par,degl}).\\

The solar neutrino problem becomes more detailed.
It can be formulated as

\noindent
1. Deficit of boron neutrinos\\
2. Deficit of the beryllium neutrinos. \\
The first one is strongly model dependent.
In fact, it may have the
astrophysical or/and nuclear physics explanation: say,
25\% decrease of the $S_{17}$ and 1\% decrease of the
central temperature (due to the plasma effects) are enough
to accommodate the Kamiokande result. The second deficit is
essentially  (solar) model independent and
it is almost impossible to explain it by reasonable variations
of parameters.

\subsection{Best fit of the data}

If nothing happens with neutrinos and the
flux at the Earth consists of the
electron neutrinos, then the data fix uniquely
values of fluxes which give the {\it best fit}~\cite{krsm}:

\noindent
1. Boron neutrino flux should be $\approx (0.35 - 0.40) \Phi_{B}^{SSM}$.\\
2. Beryllium neutrino flux as well as other fluxes of the intermediate
energies  (pep, N, O) give negligible contributions to the signals.\\
3. There is little or no suppression of the pp-flux.\\

Thus the energy dependence of the suppression
factor $P(E)$ can be represented as
$$
P(E < 0.5 ~{\rm MeV}) \equiv P_{pp} =
0.9 - 1, \  \  \
P(E = 0.7 - 1.5  {\rm MeV}) \equiv P_{Be} \sim 0 ,
$$
\begin{equation}
P(E > 7 {\rm MeV}) \equiv P_{B} = 0.4 - 1 .
\label{eq:sup}
\end{equation}
Large uncertainty of suppression in high
energy region is related to
the uncertainty in the original boron neutrino flux.
Kamiokande admits a mild distortion
of the recoil electron spectrum.
%which can be characterized by slope ....

Evidently the astrophysics  can not reproduce such a
picture~\cite{bar,bb}, ~\cite{cas} - \cite{degl}.
Typically one gets more strong suppression of the boron neutrino flux
than the beryllium neutrino flux.

(To reproduce  central values of signals one should suggest
that there is an additional flux which contributes
to the Kamiokande signal,
$\Delta \Phi_B \approx 0.09 \Phi_B^{SSM}$,
but does not contribute
to the Ar-production rate. This however implies
the conversion
of the electron neutrinos to muon or tau neutrinos).\\

The suppression profile can be strongly changed if one
admits the existence of muon or/and tau neutrino components in the
solar neutrino flux (which  already implies some
kind of neutrino transformations). In this case, especially if the
original boron neutrino flux is higher than in SSM,
$\nu_{\mu}$ and $\nu_{\tau}$ scattering on electrons
can give big (main) contribution to the Kamiokande
result and  the statement that beryllium neutrino
flux should be strongly suppressed is not true. Denoting by
$P_{Be}$ the suppression factor in the region of Be-neutrinos
(at the intermediate energies) we find the suppression factor
for the boron neutrinos needed to reproduce the Homestake result:
\begin{equation}
P_B = \frac{Q_{Ar}^{obs} - P_{Be} Q_{Ar,int}^{SSM}}{f_B Q_{Ar,B}^{SSM}},
\label{eq:pb}
\end{equation}
where $Q_{Ar}^{obs}$ is the measured Ar-production rate,
$Q_{Ar,int}^{SSM}$ and  $Q_{Ar,B}^{SSM}$ are
respectively the contributions of
fluxes of intermediate energies (Be, pep, N, O) and the
boron neutrino flux to
the Ar-production rate according to the reference SSM~\cite{bp}.
The original
boron neutrino flux which is needed to reproduce the Kamiokande
signal can be found from condition
$R_{\nu e} =
f_B [P_B + (1 - P_B)/6]$, where
$R_{\nu e} \equiv \Phi^{obs}_{B}/\Phi^{SSM}_{B}$ is the
suppression factor observed by  Kamiokande.
Substituting
$P_B$ from (\ref{eq:pb}) in this condition  one gets
\begin{equation}
f_B \approx 6R_{\nu e} - 5 \frac{Q_{Ar}^{obs}}{Q_{Ar,B}^{SSM}} +
P_{Be} \frac{Q_{Ar,int}^{SSM}}{Q_{Ar,B}^{SSM}} .
\label{eq:fbor}
\end{equation}
Then  predicted values of $Q$
\cite{bp} and the  central values of experimental signals
give according to (\ref{eq:fbor})
$f_B \approx 1.4$ for $P_{Be} = 0.5$ and
$f_B \approx 2.1$ for $P_{Be} = 1$. That is to avoid
any suppression of the Be-neutrino flux one needs two
times larger original boron neutrino flux.  For
$2\sigma$ smaller value of $R_{\nu e}$
and  $f_B =1$ the suppression as weak as $P_{Be} \sim 0.7$
becomes allowed. Thus if neutrinos undergo conversion, the
Be-neutrino flux may not be suppressed.

In the case of  weak suppression of the beryllium line the
pp-neutrino flux should be suppressed according to  the Gallium result:
\begin{equation}
P_{pp} = \frac{1}{Q_{Ge,pp}^{SSM}}
    \left[ Q_{Ge}^{obs} - P_{Be} Q_{Ge,int}^{SSM} -
P_B f_B Q_{Ge,B}^{SSM} \right],
\label{eq:ppp}
\end{equation}
where $Q_{Ge,pp}^{SSM}$ and  $Q_{Ge,int}^{SSM}$
are predicted contributions to the Ge-production rate
from pp - flux and the intermediate energy fluxes.
For $P_{Be} \sim 0.7$ one gets from (\ref{eq:ppp})
$P_{pp} \sim 0.6$.

Such a situation (weak suppression of the beryllium flux
and appreciable suppression of the pp-flux)
is realized e.g. in the case of vacuum
oscillations~\cite{calab}.

\section{Neutrino parameters and  neutrino fluxes}

Although the solar neutrino problem can
be formulated in practically
model independent way the implications
to the neutrino physics strongly
depend on the original fluxes.
There are several recent studies
of the particle physics solutions of the
solar neutrino problem according to
the (solar) model independent approach
{}~\cite{hala1,krsm,bfl,bero,krpe1,krpe2}.

\subsection{Long length vacuum oscillations}

These oscillations can reasonably well reproduce the
desired suppression. For
$\Delta m^2 > 3 \cdot 10^{-11}$ eV$^2$
the pp-neutrino flux is in the
region of averaged oscillations, where
$P = 1 - 0.5 \sin^2 2\theta$, the Beryllium neutrinos are in the
fastly oscillating part  of the $P(E)$
(so that one expects an appreciable
time variations of the Be-neutrino flux due to annual
change of distance
between the Sun and the Earth). Boron neutrinos are in the first
(high energy) minimum of $P(E)$. This allows one to reach the
inequality $P_{pp} > P_B > B_{Be}$ implied by (\ref{eq:sup}).
However,  there is an obvious relation
between maximal suppression of the Be-line and suppression of
pp-neutrinos:
$P_{Be, min} = 2 P_{pp} -1$,
and due to this the best fit configuration (\ref{eq:sup})
is  not realized.  Good  fit can be obtained
for moderate suppression of the Be-line and $\sim 0.6$ suppression of
the pp-neutrinos. The fit becomes better for increased values of
$f_B$ ~\cite{bero}.

With diminishing $f_B$
the  needed  suppression of B-neutrino flux due to the
oscillations becomes weaker. Therefore
for fixed values of $\Delta m^2$
the allowed regions of parameters
shift to smaller $\sin^2 2\theta$ ~\cite{bero,krpe1}.
In particular, for
$f_B = 0.7$,
the region is  at
$\sin^2 2\theta < 0.7$
thus satisfying the  bound from SN87A~\cite{ssb}.
For  $f_B \sim  0.4$  mixing can be as small as
$\sin^2 2\theta < 0.5 - 0.6$. Moreover, for
$f_B = 0.5$ the allowed region appears at
$\Delta m^2 \sim 5 \cdot 10^{-12}$ eV$^2$
which corresponds to
the Be-neutrino line in the first high energy minimum of $P$,
pp-neutrinos in the first maximum of the $P$ and
high energy part
of the boron neutrino spectrum out of suppression pit.
No appreciable time variations are expected.
Such a configuration is quite similar to that of
very small mixing MSW solution  which further
increases the ambiguity of situation. Distortion
of pp-neutrino spectrum is the signature of the solution~\cite{krpe2}.

Depending on neutrino parameters and
$f_B$, $f_{Be}$ ...  one can get  variety
of distortions of the boron neutrino energy
spectrum~\cite{krpe1}.

Being excluded at $f_B = 1$, the oscillations into sterile neutrino
are allowed for $f_B < 0.7$~\cite{krpe2}.\\

\subsection{Resonance flavor conversion}

It can precisely reproduce the desired energy dependence of the
suppression factor (\ref{eq:sup}). In the region of small mixing angles
one has
\begin{equation}
P_{pp} \sim 1, \ \ \  P_{Be} \sim 0, \ \ \
P_B \sim exp(-E_{na}/E),
\label{eq:small}
\end{equation}
where $E_{na} \equiv \Delta m^2 l_n \sin^2 2\theta$.
Additional contribution to Kamiokande $\Delta f_B \approx 0.09$,
follows from  scattering of the converted $\nu_{\mu}$ ($\nu_{\tau}$)
on electrons due to the neutral currents.
With diminishing $f_B$ the  suppression due
to conversion  should be relaxed,  and
therefore  $\sin^2 2\theta$ should decrease according to
(\ref{eq:small}) ~\cite{krsm,bfl}.
At  $\Delta m^2 =6  \cdot 10^{-6}$ eV$^2$
the best fit of the data for flavor mixing
corresponds to the pairs of parameters~\cite{krsm}:
$(f_B ,~ \sin^2 2\theta)$ =
$(0.4,~ 1.0 \cdot 10^{-3})$,
$(0.75,~ 4.3 \cdot 10^{-3})$,
$(1.0, ~ 6.2 \cdot 10^{-3})$,
$(1.5,~ 9 \cdot 10^{-3})$,
$(2.0,~ 10^{-2})$ .
The  decrease of $f_{Be}$
gives an additional small shift of the allowed region
to  smaller values of $\sin^2 2\theta$.
A consistent description
of the data has been found for~\cite{krsm}
$$
f_B \sim 0.4 - 2.0 .
$$

For unfixed values of the original fluxes,
$f_B$, $f_{Be}$ ...,
the allowed region of neutrino parameters  is controlled
immediately by Gallium data and by the
``double ratio". Namely,
the mass squared difference
\begin{equation}
\Delta m^2 = (6~ ^{+4}_{-3}) \cdot 10^{-6} {\rm eV^2}, \ \ \
%m = (2 - 3) \cdot 10^{-3} {\rm eV}
\end{equation}
is restricted  by Gallium results which imply
that the adiabatic edge of the suppression  pit is
in between the end point of the pp-neutrino spectrum and the Be-line.
This bound does not depend on mixing angle in a wide region of $\theta$.
(For sterile neutrinos the bound is approximately the same).
For fixed $\Delta m^2$ the  mixing $\sin^2 2\theta$ is determined by
the ``double ratio''
$$
R_{H/K} \equiv {R_{Ar} \over R_{\nu e}}\;,
$$
where
$R_{Ar} \equiv Q^{obs}_{Ar}/Q^{SSM}_{Ar}$
is the suppressions of
signals in $Cl$--$Ar$ experiment and
$Q^{SSM}_{Ar}$ is
the predictions in the reference model~\cite{bp}.
The experimental value, $R_{H/K} = 0.65 \pm 0.11$,  admits
$\sin^2 2\theta = 1.0 \cdot 10^{-3} - 1.5 \cdot 10^{-2}$ .
Similar bound exists for the conversion to
sterile neutrinos if one restricts the original boron neutrino flux by
$\Phi_B \leq  1.5 \Phi^{SSM}_B$.\\

For  very small mixing solution:
$f_B \sim 0.5$, $\sin^2 2\theta_{es} \sim 10^{-3}$,   all
the effects of
conversion in the high energy part of the boron neutrino spectrum
($E > 5 - 6$ MeV)  become very weak. In particular,
the distortion of the energy spectrum disappears, and the
ratio charged-to-neutral currents
$(CC/NC)^{exp}/(CC/NC)^{th}$ approaches 1.
Thus studying just this
part of spectrum it will
be difficult to identify the solution (e.g., to distinguish
the conversion and the astrophysical effects).
Recent calculations in SSM
with diffusion of heavy elements give larger boron neutrino
flux~\cite{prof,bah}, so that even
with $25 \%$ decrease of nuclear cross-section  and
$1\%$ decrease of central temperature of the Sun one still needs an
appreciable conversion effect. This gives a hope that
the problem can be resolved by SuperKamiokande/SNO experiments.\\

With increase of $f_B$ the fit of the data in the large mixing
domain becomes better~\cite{krsm}
. Here the Kamiokande signal can be explained
essentially by NC effect and  mixing can be relatively small.
Be- neutrino flux is sufficiently suppressed and
suppression of the pp-neutrinos is rather weak. For $f_B = 2 $ the values
$\sin^2 2\theta = 0.2 - 0.3$ become allowed. Corresponding
mass squared difference is
$\Delta m^2 = 6  \cdot 10^{-6} - 10^{-4}$ eV$^2$.\\

\subsection{The effect of third neutrino}

The analysis of data in terms of two neutrino mixing
is quite realistic, since in the most
interesting cases
(simultaneous solution of the solar and hot dark matter problems,
or solar and atmospheric neutrino problems)
third neutrino has large mass so that
its $\Delta m^2$ is beyond the resonance triangle and its mixing
to the electron neutrino is rather small. This reduces the
three neutrino task to the case of two neutrino mixing.
However, there is one interesting example where third neutrino could
influence the solutions of the solar neutrino problem.
It was considered previously~\cite{shi,asm,josh}
 and reanalyzed recently
in~\cite{ema}: The third neutrino is in the region
of the solution of the
atmospheric neutrino problem: $m_3 \sim 0.1$ eV and it has
an appreciable admixture to the
electron neutrino state. Let us represent  the $\nu_e$ as
\begin{equation}
\nu_e = \cos \phi ~ \nu' + \sin \phi ~ \nu_3
\end{equation}
where
$$
\nu' = \cos \theta ~ \nu_1 + \sin \theta ~ \nu_2
$$
and $\phi$ is not small.
In the case $m_3 \gg m_2$
the third neutrino $\nu_3$ ``decouples"
from the system
(as far as we deal with the Sun) and
its effect is reduced just to the averaged vacuum oscillations.
In turn  $\nu'$  converts resonantly to its
orthogonal state.
So that the survival probability can be written as~\cite{asm}
\begin{equation}
P = \cos^4 \phi ~ P_2 + \sin^4 \phi,
\label{eq:prob}
\end{equation}
where $P_2$ is two neutrino survival probability.
Additional regions of the neutrino parameters
$\Delta m^2 = (10^{-5} - 10^{-6})$ eV$^2$ and
$\sin^2 2\theta = 3\cdot 10^{-4} - 3\cdot 10^{-3}$ are allowed for
$\cos^4 \phi \sim 0.5 - 0.7$. Now  both pp- and Be- neutrinos
can be outside the $2\nu$ - suppression pit~\cite{asm},
where $P_2 \approx 1$
and according to (\ref{eq:prob})  the suppression factor for them
is ($\cos^4 \phi + \sin^4 \phi$). This allows one to get
about 1/2 suppression of the gallium production rate,
and reconcile  the
Homestake and the Kamiokande results at $2\sigma$ level.
In~\cite{ema} it is claimed that even the adiabatic solution
(when the high energy part of the boron neutrino spectrum is
on the adiabatic edge) is not excluded. Indeed, now the distortion of the
boron neutrino spectrum is weakened by factor $\cos^4 \phi$
in comparison with two neutrino case. But even this is disfavored
by the data. Large mixing with electron neutrinos
(now $\sin^2 2\phi \sim 0.75$)
is practically excluded by reactor experiment~\cite{bil}
and CHOOZ will finally check this possibility.

\subsection{On the spin-flip effects}

Resonance spin-flip precession can precisely reproduce the
suppression profile~\cite{akh,lim,kras},
i.e. give very good description of
averaged signals.

For values of the magnetic moment at the
upper bound:  $\mu \sim 3 \cdot 10^{-12} \mu_B$, where $\mu_B$ is the
Bohr magneton,  the strength of the magnetic field
as big as $10^6$ Gauss is needed. Traditional objection is that
this field is much stronger than usually expected one.
There is another objection. In the most of calculations
it was suggested that there is no
{\it latitude} dependence of the field which  is certainly incorrect:
the toroidal field has different polarity in the southern and
northern semispheres and there is the equatorial gap of the field.
One can think that existing calculations correspond  to some   average
field. However this means that there are regions with
even stronger  field than that mentioned above. Moreover, since the
spin-flip effect is non-linear in the field one should
calculate first the probabilities for different
latitudes and then perform the averaging over the latitude
rather than use the  average field.
In fact,  it was shown~\cite{kras} that for reasonable latitude
distributions of the field the average suppression is
too weak, moreover one expects an appreciable seasonal
effect.

Time variations of signals are  generic features  of
this solution. Where are these variations?

\section{Standard and non-standard}

The solution of the solar neutrino problem  can be reconciled  with
solutions of (all ?) other neutrino anomalies like
deficit of the atmospheric $\nu_{\mu}$- flux,
possible signal of the $\bar{\nu}_{\mu} - \bar{\nu}_{\tau}$ oscillations,
existence of  hot component of dark matter.
A number of schemes of the neutrino masses and mixing has been
suggested in this context.
Thinking in term ``standard" and ``non-standard" one can
arrive at the following (at least the most popular) scenario.

\subsection{``Standard" scenario of neutrino masses and mixing}

(i). Neutrino masses are generated by the see-saw
mechanism with
masses of the RH components $M_R = 10^{11} - 10^{13}$ GeV.
The mass scale $10^{13}$ GeV can originate, e.g., from
Grand Unification scale,
$M_{GU}$, and the Planck scale, $M_{P}$,
as $M_R \sim M_{GU}^2/M_{P}$.\\
(ii) Second mass, $m_2$,  is in the range
\begin{equation}
m_2 = (2 - 3)\cdot 10^{-3} {\rm eV},
\end{equation}
so that the
resonance flavor conversion $\nu_e \to \nu_{\mu}$
solves the solar neutrino problem. The desired
mixing angle is consistent with
\begin{equation}
\theta_{e\mu} = \sqrt{\frac{m_e}{m_{\mu}}} - e^{i \phi} \theta_{\nu},
\end{equation}
where $\theta_{\nu}$ comes from diagonalization of  neutrino mass
matrix.  This relation is similar to  corresponding relation in quark
sector which testifies for certain quark-lepton symmetry (unification). \\
(iii) The third neutrino (for $m^D \sim 100$ GeV and
$M \sim 3 \cdot 10^{12}$ GeV)
has the mass about 5 eV.
It composes the desired hot component of the dark
matter.\\
(iv) The decays of the RH neutrinos with mass $10^{12}$ GeV can
produce the lepton asymmetry of the Universe
which can be  transformed by sphalerons in to the
baryon asymmetry~\cite{fuk}.\\
(v) Large Yukawa coupling of neutrino from  the third generation,
e.g. $Y_{\nu} \sim Y_{top}$,
gives appreciable renormalization effects in the region of momenta
$M_R - M_{GU}$.
In particular, the $b - \tau$ mass ratio increases by
$(10 - 15) \%$
in the MSSM. In turn this disfavors the $b - \tau$ mass unification
for low values of $\tan \beta$~\cite{vis}.\\
(vi) Simplest schemes with  quark - lepton symmetry
lead to  mixing angle for the $e$ and $\tau$ generations:
$\theta_{e \tau} \sim (0.3 - 3) V_{td}$ which
is close to the bound from the nucleosynthesis of heavy elements
(r-processes) in the inner parts of the supernovae:
$\sin^2 2\theta_{e \tau} < 10^{-5}$  ($m_3 > 2$ eV)~\cite{qian}.\\
(vii) For $\mu -\tau$  mixing one expects~\cite{dhr}
$
\theta_{\mu \tau} \sim k V_{cb} \eta,
$
where $k = 1/3 - 3$ and $\eta \sim 0.6 - 0.7$ is the
renormalization factor.
If $m_3 > 3$ eV some part of expected region of mixing angles
is already excluded by FNAL 531.
Large part of the region can be studied by CHORUS and NOMAD. The rest
(especially $m_3 < 2$ eV) could  be covered by E 803.\\
(viii) The depth of $\bar{\nu}_{\mu} - \bar{\nu}_e$
oscillations with $\Delta m^2 \approx m_3^2$  equals
$4|U_{3\mu}|^2 |U_{3e}|^2 \approx
4|\theta _{e \tau}|^2 |\theta_{\mu \tau}|^2$. The existing
experimental bounds on
$\theta _{e \tau}$ and $\theta_{\mu \tau}$
give the upper bound on this depth:
$ < 10^{-3}$~\cite{babu}
which is too small to explain the LSND result.\\

The standard scenario does not solve the atmospheric neutrino problem.
One can consider the scheme with three degenerate neutrinos or
sacrifice the HDM suggesting that some other particles
are responsible for the structure formation in the Universe.
In the latter case $m_3 \sim 0.1$ eV and strong
$\mu - \tau$ mixing  explain via
$\nu_{\mu} - \nu_{\tau}$ oscillations  the
atmospheric neutrino deficit. Strong
$\mu - \tau$ mixing, could be related to  relatively small mass splitting
between $m_2$ and $m_3$ which implies the enhancement of the
mixing  in the neutrino Dirac mass matrix~\cite{fty}.
It could be related to  the see-saw enhancement mechanism~\cite{sss,tan}
endowed by renormalization group
enhancement~\cite{tan} or with strong mixing in the charge lepton sector
{}~\cite{leon}. \\

%All three anomalies can be explained  if neutrinos have strongly
%degenerate  mass spectrum $m_1 \approx m_2 \approx m_3$, so that the
%mass matrix has the form
%\begin{equation}
%m = m_0 I + \delta m ,
%\end{equation}
%where $\delta m \ll m_0 \approx 1 - 2$ eV.
%Such a scenario can be realized in the unique see-saw mechanism with non
%zero element $m_{LL}$. Main contribution originates with interaction
%which respects some horizontal symmetry like
%$SU(2)$ or $S_4$ or permutation symmetry. Nontrivial observation is that
%the desired mass splitting $\delta m$ can be generated by the
%standard see-saw contribution.
%
%The scenario has however the potential problem with
%neutrinoless double beta decay.
%In the simplest version the effective Majorana mass of the electron
%neutrino is just $m_0$ which exceeds the present bound
%$m_e < 0.7 - 1.5 $ eV.
%To satisfy this bound one should admit the cancellation in the
%effective mass which implies large mixing of the electron neutrino
%with other component and further complication of the model.\\

\subsection{More neutrino states?}

Safe way to accommodate all the anomalies is to introduce
new neutrino state.
As follows from LEP bound on the
number of neutrino species this state
should be sterile (singlet of standard group).
Taking  into account also
strong bound on  parameters of oscillations into sterile neutrino from
Primordial Nucleosynthesis one
can write the following
``scenario"~\cite{cald} $^-$~\cite{maro}.\\

\noindent
(i) Sterile neutrino has the mass $m_S \sim (2 - 3) \cdot 10^{-3}$ eV
and mixes with $\nu_e$, so that the resonance conversion
$\nu_e -\nu_s$ solves the solar neutrino problem;\\
(ii) Masses of  ${\nu}_{\mu}$ and  ${\nu}_{\tau}$
are in the range 2 - 3 eV,  they supply
the  hot component of the DM; \\
(iii)  ${\nu}_{\mu}$ and  ${\nu}_{\tau}$
form the pseudo Dirac neutrino with large (maximal) mixing and
the oscillations ${\nu}_{\mu} - {\nu}_{\tau}$
explain the atmospheric neutrino problem;\\
(iv) $\nu_e$ is very light:  $m_1 < 2\cdot10^{-3}$ eV. The
$\bar{\nu}_{\mu} - \bar{\nu}_e$ mixing can be strong enough
to explain the LSND result.\\
(v) However production of heavy elements in
supernova via  ``r-processes" is problematic for this scenario.\\

Sterile neutrino can be used
to explain the atmospheric
neutrino problem in the context of standard scenario,
if one ignores the Nucleosynthesis bound.

\subsection{Sterile neutrino or light singlet fermion?}

Introducing sterile neutrino $S$ one encounters several questions:\\
\noindent
What is the origin of this neutrino?\\
How it mixes with usual neutrinos?\\
How one can explain its small mass? \\

1. {\it Origin.}  Natural candidate is
if course, the RH neutrino component.
However in this case  the see-saw mechanism does not operate.
Then $S$  could be the component of the multiplet of extended gauge
symmetry - like  $SO(10)$- singlet from  27-plet of $E_6$
\cite{ma}.
In~\cite{bermo} it was suggested that $S$ is the mirror neutrino
from mirror standard model.
In all these cases one has three singlet fermions.

Let us consider another possibility~\cite{cjs1,cjs2}:\\
(i). $S$ has an origin beyond usual
fermionic structure, and in particular,
beyond the see-saw mechanism. So that the see-saw explains
the lightness of the active  neutrinos in the usual way. \\
(ii). $S$ has no generation structure and probably is generation blind.
There is only one light singlet fermion (although
this is not necessarily).\\
(iii). Supersymmetry may be a natural framework of the appearance of such a
fermion. A number of singlet superfields was
introduced for different purposes: to generate $\mu$ term,
to realize PQ-symmetry breaking, to break spontaneously lepton
number, etc..   String theory typically  supplies some
singlets. Fermionic components of these superfield could be
identified with desired sterile neutrino.\\

2. {\it Mixing}. The standard see-saw structure
\begin{equation}
h L \nu^c H_2 +  M \nu^c \nu^c
\label{eq:sees}
\end{equation}
involves three fields: doublet neutrino, RH neutrino component,
$\nu^c$, and Higgs doublet, $H_2$. Correspondingly,  there are three
possible ways to mix S with active neutrinos:

(i) via direct coupling to the left handed neutrinos:
\begin{equation}
\epsilon L S  H_2 .
\end{equation}
The parameter $\epsilon$ is of the order
\begin{equation}
\epsilon  \sim \frac{m_{3/2}}{M_P},
\end{equation}
where $M_P$ is the Planck mass. This means that $S$ could
be the field from hidden sector which is mixed with usual neutrinos via
gravitational interactions.

(ii). mixing via interactions with RH neutrinos:
\begin{equation}
\lambda \nu^c S y,
\label{eq:rh}
\end{equation}
where $y$ is an additional singlet field which acquires the VEV
$<y> \sim m_{3/2}$ as the result of SUSY breaking.
The interactions (\ref{eq:sees}) and
(\ref{eq:rh}) allow one  to explain simultaneously both
the mixing of $S$ with neutrinos and the desired mass of $S$
without introduction of new mass scales~\cite{cjs1,maro}.

(iii). mixing via Higgsino:
\begin{equation}
\frac{\mu}{<S>} H_1 H_2 S   +
\epsilon' L H_2.
\end{equation}
The first term which  mixes  S with Higgsino
can be responsible also for generation of the
$\mu$-term . Second term mixes Higgsino with neutrino
thus breaking R-parity. It can be generated spontaneously by
the interaction (\ref{eq:sees}), if sneutrino acquires non-zero VEV,
or explicitly by, e.g., gravitational interactions.\\

3. {\it Mass}.
Spontaneous violation of global symmetry, ( $U(1)_G$ in the
simplest case)  like Peccei-Quinn or lepton
number symmetries or  horizontal symmetry
leads to  appearance of the massless boson.
In the limit of exact supersymmetry the fermion partner is also
massless. However,  violation of supersymmetry results
in  generation of mass of $S$. In supergravity
one has typically $m_S \sim m_{3/2}$.
The mass  $m_S$ can be further suppressed
by special choice of (i) the superpotential
or (ii) K\"ahler potential. In the first case it is quite easy to get
$m_S \sim  m_{3/2}^2/M_G$ which leads to desired value of $m_S$
for $M_G \sim 10^{16}$ GeV ~\cite{cjs2}.

Using  non-minimal kinetic
terms one can suppress $m_S$ at tree level, one loop or
even two loops. In the case of strict
no-scale supergravity the mass is generated in three loops
which is
sufficient to explain the smallness of $m_S$ for rather natural
values of parameters~\cite{cjs2}.

Another possibility~\cite{cjs1} is to use
$R$-symmetry to  protect the mass, to forbid undesired
mixing of $S$ and to ensure that $S$ does not acquires a VEV.
(In this case $R$-parity can be conserved).
For more details see~\cite{chun}.\\

Thus discovery of solar neutrinos conversion
into sterile neutrinos
(and future experiments will be able to do this )
may give the hint to supersymmetry  and to really very
rich physics beyond the standard model.

\section{Conclusion}

Before Gallium experiments we had the following criteria:
The counting rate much below 70 - 75 SNU is the proof of new
neutrino physics (oscillations, conversion etc.).
The counting rate much bigger 75 SNU testifies for
astrophysical solution. Nature has chosen precisely 75 SNU,
and has stayed us in uncertain situation for more than 5 years.
Although present rather precise data
strongly indicate new physics we have not passed through
simple {\it a priori} criteria.

Now we have new chance. {\it A priori} criteria are:
distortion of the energy spectrum of boron neutrinos,
anomalous ratio of charged to neutral current number  events,
day-night effect, seasonal variations...

Following  previous logic of Nature one can
imagine that neither distortion nor
time variations or CC/NC anomaly will be observed.
By the way, this is quite possible situation,
e.g. if the original boron neutrino flux is small, and
the transitions of this flux is not needed.
Is this the proof of Astrophysical solution?
For boron neutrinos - Yes.
Then to explain other experiments one
should suggest strong suppression
of the Beryllium neutrino flux.
This is realized, e.g.,  in very small mixing MSW - solution.
To proof this one should wait again - wait for
experiments which are sensitive to  beryllium neutrinos
and  in this case BOREXINO results can be decisive...

Will Nature again play with us?
Let us see what will happen ...

%\section*{Acknowledgments}
%I would like to thank ......

\end{document}